\documentclass[aps,twocolumn,prb,showpacs,floatfix,amsmath,amssymb,superscriptaddress]{revtex4-1}
\usepackage{graphicx}
\usepackage{bm}
\usepackage{hyperref}
\usepackage{color}
\usepackage{natbib}
\usepackage{changes,cancel,amsmath, amssymb}
\usepackage{braket,float,soul}

\newcommand{\tmmathbf}[1]{\ensuremath{\boldsymbol{#1}}}
\newcommand{\tmop}[1]{\ensuremath{\operatorname{#1}}}

\DeclareMathOperator{\tr}{\mathrm{Tr}}

\graphicspath{{../Draft/} }
\begin{document}
\title{Feasible model for photo-induced interband pairing}

\author{S. Porta*}
\affiliation{Dipartimento di Fisica, Universit\`a di Genova, 16146 Genova, Italy}
\affiliation{SPIN-CNR, 16146 Genova, Italy}

\author{L. Privitera*}
\affiliation{Institute for Theoretical Physics and Astrophysics, University of W\"urzburg, 97074 W\"urzburg, Germany}

\author{N. Traverso Ziani}
\affiliation{Dipartimento di Fisica, Universit\`a di Genova, 16146 Genova, Italy}
\affiliation{SPIN-CNR, 16146 Genova, Italy}
\affiliation{Institute for Theoretical Physics and Astrophysics, University of W\"urzburg, 97074 W\"urzburg, Germany}

\author{M. Sassetti}
\affiliation{Dipartimento di Fisica, Universit\`a di Genova, 16146 Genova, Italy}
\affiliation{SPIN-CNR, 16146 Genova, Italy}

\author{F. Cavaliere}
\affiliation{Dipartimento di Fisica, Universit\`a di Genova, 16146 Genova, Italy}
\affiliation{SPIN-CNR, 16146 Genova, Italy}

\author{B. Trauzettel}
\affiliation{Institute for Theoretical Physics and Astrophysics, University of W\"urzburg, 97074 W\"urzburg, Germany}
\affiliation{W\"urzburg-Dresden Cluster of Excellence ct.qmat, Germany}

\begin{abstract}
Recently, it has been proposed that a mechanism for the appearance of non-equilibrium superconductivity in a resonantly driven semiconductor with repulsive interband interactions exists.~\cite{Goldstein_PRB15} The underlying microscopic model relies on the appearance of a specific fermionic dissipation mechanism and the careful simultaneous tailoring  of the electronic dispersion relation and electron-electron interactions.
We, instead, show that the phenomenon is rather general and does not need a special fine tuning of parameters. By considering a pair of bands with locally the same sign of concavity, we demonstrate that interband pairing arises under the natural assumption of the presence of phononic baths and radiative recombination. In light of these findings, we demonstrate how the emergence of superconductivity can be understood in terms of standard equilibrium interband BCS theory.
\end{abstract}

\maketitle
\section{Introduction}
The time-dependent manipulation of quantum systems represents a promising way towards the dynamic realization of quantum phases of matter. Sudden parameter variations are useful for inspecting quantum thermalization~\cite{Cardy2006,Eisert2015,dalessio2016,Mitra2018,Porta2016,Porta20181}; short electromagnetic pulses can, for example, induce phase transitions~\cite{transition} and drive higher harmonic generation~\cite{Ghimire_Nphys18}. Moreover, periodic drivings allow to induce topological band structures and boundary states~\cite{Lindner2011,Oka_PRB11,Jiang_PRL11,Jotzu2014,Goldman2016,Privitera,mciver2018light}, and to create new phases of matter, such as time-crystals~\cite{Sacha_2017,Watanabe,Else2016,Gambetta}.\\
An important sub-field  of time-dependent quantum engineering in solids deals with  the control of the superconducting order parameter. The importance of dynamic perturbations on the superconducting order parameter has been known for a long time~\cite{Wyatt,Dayem,Eliashberg}. Recently, however, the technological progress in the generation of intense sub-picosecond laser pulses, marked a renewed interest in the field~\cite{Cavalleri}. Striking signatures of transient out of equilibrium superconductivity have been observed in cuprates and doped fullerenes: the ultimate aim is to engineer room temperature superconductivity, although serious limitations like heating still need to be overcome. All these observations have been associated to the action of the laser on lattice degrees of freedom, in particular optical phonons, e.g to light-enhanced electron-phonon coupling. In the context of doped fullerenes, a possible electronic mechanism for intralevel pairing has been conjectured~\cite{Nava_Nphys17}.\\
%
A novel scheme for generating \textit{interband} superconducting pairing in periodically driven semiconductors has been proposed~\cite{Goldstein_PRB15,Hart_arxiv18}. In this case, however, the laser couples to the electronic degrees of freedom, and the crucial interplay of driving and dissipation leads to a steady state characterized by superconducting correlations.\\
The original proposals for  such a state relied on particular fermionic dissipative baths able to exchange particles with the two bands of the semiconductor involved in the interband pairing. Moreover, a finite value of the order parameter  in the steady state required the concurrent tuning of band dispersion and electronic interactions. \\
In our work, we develop a more realistic model for steady state interband superconductivity. We consider a two-band semiconductor resonantly driven by a laser, and we include two physically relevant intraband relaxation processes namely acoustic phonons and radiative recombination. The steady state reached by the system, for which we provide a phase diagram, can develop interband superconducting correlations. The required conditions are strong laser amplitudes and strong repulsive electron-electron interactions. Moreover, we elucidate the role of the various relaxation processes by inspecting the transient dynamics. We finally devote a section to the discussion of a Bogoliubov-de Gennes effective model that allows us to qualitatively address the physics of the steady state in terms of time-independent interband pairing~\cite{Moreo_PRB09}.\\
The structure of the article is as follows: In Sec.II, we introduce the model and explain how we treat its dynamics; in Sec.III, we describe our results and discuss a simplified model. Finally, in Sec.IV, we summarize our results and draw conclusions.
\section{Model and methods}
\subsection{Model}
\begin{figure}[!ht]
	\centering
	\includegraphics[width=0.8\linewidth]{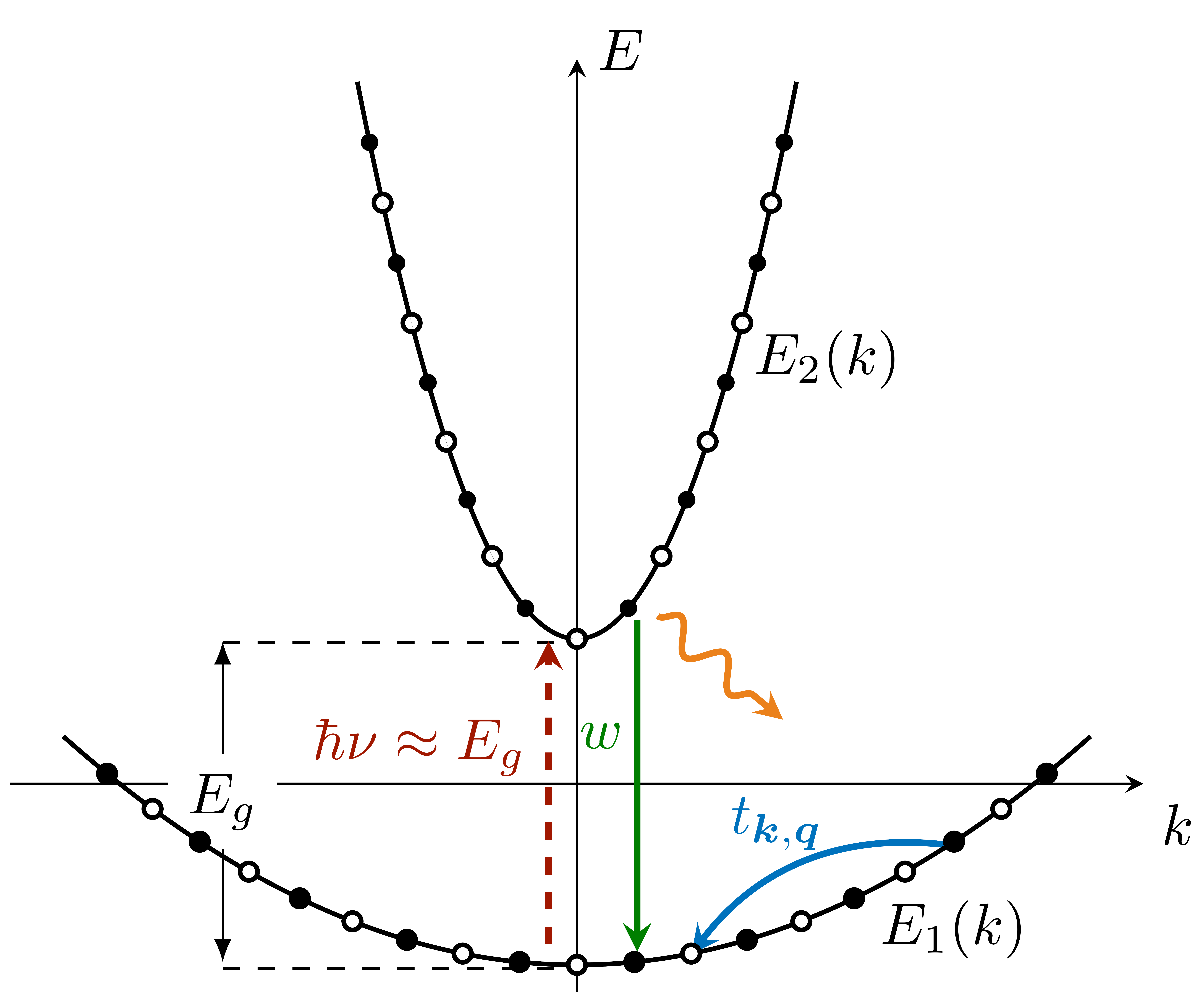}
	\caption{Sketch of a restricted momentum region of the two bands in presence of a laser (red dashed arrow) tuned to a frequency $ \nu $ equal to the amplitude of the band distance $ E_{g} $ at $ \tmmathbf{k}=0 $. Emission of intraband acoustic phonons (blue curved arrow) and radiative recombination (green  arrow) accompanied by photon emissions (orange wavy arrow) are indicated.}
	\label{fig:model}
\end{figure}
We consider a $d$-dimensional semiconductor with only two non-degenerate bands close to the Fermi energy, namely, the valence ($ \alpha=1 $) and conduction ($\alpha=2$) bands. These are coupled by means of a laser light, whose frequency is tuned at resonance around a single point in the Brillouin zone (BZ), where the band distance is $E_g$. At the same time, a superconducting pairing, whose \textbf{amplitude} is computed self-consistently by taking electron-electron interactions into account, is allowed. The system is described by the Hamiltonian $ H_{\mathrm{sys}}=H_0+H_{\mathrm{int}} $, where ($ \hbar=1 $, lattice constant $ a=1 $) 
\begin{align}
&H_{0} = \sum_{\tmmathbf{k}, \alpha} E_\alpha(\tmmathbf{k})
c^{\alpha\dag}_{\tmmathbf{k}}\label{eq:Hsys} c_{\tmmathbf{k}}^{\alpha}+\Omega(t)\sum_{\tmmathbf{k}, \alpha,\beta}c^{\alpha\dag}_{\tmmathbf{k}}\sigma^x_{\alpha\beta}c^{\beta}_{\tmmathbf{k}}\; ,\\
&
H_{\mathrm{int}} = \frac{i}{2}\Delta\sum_{\tmmathbf{k}, \alpha,\beta}c^{\alpha\dag}_{\tmmathbf{k}}\sigma^y_{\alpha\beta}c^{\beta\dag}_{-\tmmathbf{k}}-\frac{i}{2}\Delta^*\sum_{\tmmathbf{k}, \alpha,\beta}c^{\alpha}_{\tmmathbf{k}}\sigma^y_{\alpha \beta}c^{\beta}_{-\tmmathbf{k}}\ .
\end{align}
Here $ c^{\alpha}_{\tmmathbf{k}} $ is the fermionic annihilation operator in the $ \alpha $-band with momentum $ \tmmathbf{k} $, $ E_\alpha(\tmmathbf{k})$ is the $ \alpha $-band  dispersion relation, $ \Omega(t)=\Omega \cos(\nu t) $ is set by the laser frequency $ \nu $ and the Rabi frequency $ \Omega $.  Moreover, the complex order parameter $ \Delta $ quantifies the Cooper pairing between electrons in different bands and is defined as $ \Delta^*=\frac{V}{N} \sum_{\tmmathbf{k}} \braket{c^{2\dag}_{\tmmathbf{k}}c^{1\dag}_{-\tmmathbf{k}}}  $, where $V$ parametrizes the strength of electron-electron interactions and $ N $ is the number of electrons in the system. The relevant part of the dynamics takes place in a small region of $k$-space around the resonance point that we conventionally fix to be $\tmmathbf{k}=0$ such that we take $\left| \tmmathbf{k}\right| \leq\delta_k$. For this reason, we can assume $V(\tmmathbf{k})$ to be $k$-independent. It is worth to note that, differently from $\Delta$, the parameter $V$ does not have to be computed self-consistently, since it parametrizes the bare, repulsive, electron-electron interaction. Moreover, again due to the fact that the physics takes place in a small region in $k$-space, we expand the dispersion relation up to second order~\cite{Note_gap}. For simplicity we postulate spherical symmetry, ${E_\alpha(\tmmathbf{k})= A_\alpha \tmmathbf{k}^2 +(-1)^\alpha E_g/2}$. For technical reasons that will be explained below, we consider the scenario where the conduction and valence bands have same effective mass sign, i.e. $ A_1, A_2>0 $, in the BZ window where the laser is resonant.
Furthermore, in order to get a realistic picture of the system (see Fig.~\ref{fig:model}), we study the effect of two relevant bosonic baths, namely the phonons and the radiative field.

As far as phonons are concerned, we consider the acoustic branch, described by the Hamiltonian $
{H^{\mathrm{ph}}_{\mathrm{bath}} = \sum_{\tmmathbf{q}} v |\tmmathbf{q}|
a^{\dag}_{\tmmathbf{q}} a_{\tmmathbf{q}}^{}} 
$, where $ a^\dag_{\tmmathbf{q}} $ creates a phonon with momentum $ \tmmathbf{q} $ and $ v $ is their constant and isotropic velocity. The electron-phonon coupling is modelled by the Fr\"ohlich Hamiltonian~\cite{Ridley_book}
 \begin{equation}\label{eq:Frohlich}
H_{\mathrm{sys} - \mathrm{bath}}^{\mathrm{ph}} = \sum_{\tmmathbf{k}, \tmmathbf{q}, \alpha}
t_{\tmmathbf{k},\tmmathbf{q}} (c_{\tmmathbf{k}+\tmmathbf{q}}^{\alpha \dag}
c^{\alpha}_{\tmmathbf{k}} a^{\dag}_{-\tmmathbf{q}} + h.c.)\ ,
\end{equation}
where $ t_{\tmmathbf{k},\tmmathbf{q}} $ represents the momentum dependent coupling strength. We assume that phonons can induce transitions only within the same band, since interband phonons are typically suppressed~\cite{Esin_PRB18,Tandon_JAP15} due to symmetry reasons~\cite{Dyakonov_book}. Note that, in principle, we can consider additional optical branches without significantly affecting our results, as long as the coupling between the optical phonons and the laser is negligible.\\
We also take into account the possibility of interband radiative recombination processes~\cite{Esin_PRB18}, where a conduction band electron relaxes to the valence band and emits a photon.  In this case, as opposed to the phononic bath, the emission is associated to a pseudospin-flip to obey the angular momentum selection rules and can then only take place between different bands. The corresponding contribution to the Hamiltonian is $ {H_{\mathrm{sys} - \mathrm{bath}}^{\mathrm{rr}} = \sum_{\tmmathbf{k}}
w (c_{\tmmathbf{k}}^{1 \dag}
c^{2}_{\tmmathbf{k}} b^{\dag}_{\tmmathbf{0}} + h.c.)} $, where $ b^\dag_{\tmmathbf{0}} $ creates a photon with energy $ \omega_0(\tmmathbf{k}) = E_2(\tmmathbf{k})-E_1(\tmmathbf{k}) $ and $ w $ is the coupling intensity.\\
\subsection{Methods}
In order to study the dynamics  after the laser is switched on, we define the reduced density matrix $\rho_{\mathrm{sys}} =
\mathrm{Tr}_{\mathrm{bath}} \{ \rho_{\mathrm{tot}} \}$. Here, $\rho_{\mathrm{tot}}$ is the density matrix of the full system. The expectation value $\langle O \rangle (t) $ of an observable $O$ involving only the electronic degrees of freedom can be computed as $\langle O \rangle (t)= \tr\left\lbrace \rho_{\mathrm{sys}}(t) O \right\rbrace$. Since the exact form of $\langle O \rangle (t) $ is in general not available,  usually  approximate equations of motion are employed~\cite{Puri_book}. Here, the simplifications arise from the fact that the driving is tuned to resonance with the relevant energy gap of the system ($\nu=E_g$), allowing for the Rotating wave approximation (RWA), and from the assumption of perfectly Markovian bosonic baths. Moreover, we treat the system-baths coupling perturbatively, up to second order. We further neglect Lamb shift corrections because they only slightly renormalize the parameters~\cite{Haug_book,Schlosshauer_book}. Within these approximations, the evolution equation of the reduced density matrix is of the standard Lindblad form~\cite{Puri_book,Haug_book}
\begin{equation}\label{eq:dtrho}
\frac{d}{d t} \rho_{\mathrm{sys}} = - i [H_{\mathrm{sys}}, \rho_{\mathrm{sys}}]
+\mathcal{L}_{\mathrm{sys}} \rho_{\mathrm{sys}}\; ,
\end{equation}
where the Lindbladian acts on the reduced density matrix as
\begin{widetext}
 \begin{equation}
\mathcal{L}_{\tmop{sys}} \rho_{\tmop{sys}} = \sum_{\tmmathbf{k},
	\tmmathbf{q}, \tmmathbf{k}',
	\tmmathbf{q}'}\sum_{ \alpha, \beta} \left[ \left(\mathcal{S}^{\alpha \dag}_{\tmmathbf{k},
	\tmmathbf{q}} \rho_{\tmop{sys}} \mathcal{S}^{\beta}_{\tmmathbf{k}',
	\tmmathbf{q}'} - \rho_{\tmop{sys}} \mathcal{S}^{\beta}_{\tmmathbf{k}',
	\tmmathbf{q}'} \mathcal{S}^{\alpha \dag}_{\tmmathbf{k}, \tmmathbf{q}}\right)
W^{ \beta(1)}_{\tmmathbf{q},\tmmathbf{k}', \tmmathbf{q}'} + \left(\mathcal{S}^{\alpha}_{\tmmathbf{k},
	\tmmathbf{q}} \rho_{\tmop{sys}} \mathcal{S}^{\beta \dag}_{\tmmathbf{k}', \tmmathbf{q}'} - \rho_{\tmop{sys}} \mathcal{S}^{\beta \dag}_{\tmmathbf{k}',
	\tmmathbf{q}'} \mathcal{S}^{\alpha}_{\tmmathbf{k}, \tmmathbf{q}}\right)
W^{\beta(2)}_{ \tmmathbf{q},\tmmathbf{k}', \tmmathbf{q}'} + h.c.\right] \;.,
\end{equation}
\end{widetext} 
with
\begin{equation}
\label{eq:defS}
\mathcal{S}^{\alpha}_{\tmmathbf{k}, \tmmathbf{q}} =
c^{\alpha \dag}_{\tmmathbf{k}}c_{\tmmathbf{k}+\tmmathbf{q}}^{\alpha} \,
\end{equation}
and  $W^{\alpha\beta(1)}_{\tmmathbf{q},\tmmathbf{k}', \tmmathbf{q}'} =  \Gamma_{\tmmathbf{k}',\tmmathbf{q}}^{\alpha\beta}\left[1 + n_B(\tmmathbf{q}) \right] \delta(\tmmathbf{q}-\tmmathbf{q}')$,
$W^{\alpha\beta(2)}_{\tmmathbf{q},\tmmathbf{k}', \tmmathbf{q}'} =  \Gamma_{\tmmathbf{k}',\tmmathbf{q}}^{\alpha\beta}\ n_B(\tmmathbf{q})
\ \delta(\tmmathbf{q}-\tmmathbf{q}')$;  $n_B(\tmmathbf{q})$ is the occupation number of the bosonic mode with momentum $\tmmathbf{q}$. The dissipation rates are
\begin{equation}
\Gamma_{\tmmathbf{k}',\tmmathbf{q}}^{\alpha\beta} = \pi | t_{\tmmathbf{k}',\tmmathbf{q}}  |^2\
\delta \left[E_{\beta}
(\tmmathbf{k}')-E_\beta(\tmmathbf{k}'+\tmmathbf{q})-v|\tmmathbf{q}|\right] \delta_{\alpha, \beta} 
\label{eq:gammaph}
\end{equation}
for the phononic bath and 
\begin{equation}
\Gamma_{\tmmathbf{k}',\tmmathbf{q}}^{\alpha\beta} = \pi |w|^2 \delta_{\alpha, 3-\alpha}
\end{equation}
for the photons. It is convenient to introduce here the important quantities
\begin{equation}
\label{eq:Gph}
\Gamma^\mathrm{ph}=\pi | t_{\tmmathbf{k}',\tmmathbf{q}}  |^2\
\delta \left[E_{\beta}
(\tmmathbf{k}')-E_\beta(\tmmathbf{k}'+\tmmathbf{q})-v|\tmmathbf{q}|\right]\,,
\end{equation} 
\begin{equation}
\label{eq:Grr}
\Gamma^\mathrm{rr}=\pi |w|^2\,.
\end{equation} 
Throughout the article, we will assume that $\Gamma^{\mathrm{ph}}$ is essentially constant~\cite{EPrates} -- see Appendix B. On the basis of Eq.~\eqref{eq:dtrho}, we can derive the equation of motion 
\begin{widetext}
		\begin{equation}\label{eq:dto}
\begin{split}		
\frac{d}{d t} \langle O \rangle = &- i \langle
\left[O, H_{\tmop{sys}}\right] \rangle + \\ &\sum_{\tmmathbf{k},\tmmathbf{k}',
	\tmmathbf{q}} \sum_{\alpha,\beta}\Gamma_{\tmmathbf{k}',\tmmathbf{q}}^{\alpha\beta}  \left\{ [1 + n_B\tmmathbf{q}
]_{} \langle \mathcal{S}^{\beta}_{\tmmathbf{k}',
	\tmmathbf{q}} \left[ O,  \mathcal{S}^{\alpha \dag}_{\tmmathbf{k},
	\tmmathbf{q}}\right] +
\left[\mathcal{S}^{\alpha}_{\tmmathbf{k},\tmmathbf{q}}, O\right]\mathcal{S}^{\beta \dag}_{\tmmathbf{k}', \tmmathbf{q}}
\rangle + \right. \left. n_B\tmmathbf{q}  \langle \mathcal{S}^{\beta\dag}_{\tmmathbf{k}',
	\tmmathbf{q}} \left[ O,  \mathcal{S}^{\alpha }_{\tmmathbf{k},
	\tmmathbf{q}}\right] +
\left[\mathcal{S}^{\alpha\dag}_{\tmmathbf{k},\tmmathbf{q}}, O\right]\mathcal{S}^{\beta }_{\tmmathbf{k}', \tmmathbf{q}}
\rangle \right\} \;.
\end{split}
	\end{equation}
\end{widetext} 
Specifically, we are interested in the evolution of the anomalous interband correlator
$ s^{21}_k=\braket{c^{2\dag}_k c^{1\dag}_k} $. Its equation of motion, in turn, involves other one-body correlators, namely the populations of valence $ n^{11}_k=\braket{c^{1\dag}_k c^1_k} $ and conduction $ n^{22}_k=\braket{c^{2\dag}_k c^2_k} $ bands, and the ordinary interband correlator $ n^{21}_k=\braket{c^{2\dag}_k c^1_k} $. The equations of motion for $n_k^{11}$, $n_k^{22}$, $n_k^{12}$ and $s_k^{12}$ form a closed set of coupled ordinary differential equations -- see appendix A -- that we solve numerically. The explicit form of the system of equations that we solve is given in Appendix A. The integration is performed by means of $ 4^{th} $ order Runge-Kutta adaptive method with an absolute accuracy of the solution of $10^{-9}$. Furthermore we discretized $k$-space with a step of $2\cdot 10^{-3}$, after checking the stability of our results with respect to this value. In addition, thanks to the spherical dispersion relation of the electrons and the baths we restrict ourselves to an effective $1$-dimensional model. We do not expect this assumption to change  the steady state behaviour of the electronic populations. Indeed, as we will see, the main effect of phonons is to shift the electrons towards the bottom of the bands, a fact that is well captured by an effective one-dimensional model.  Hereafter, we set $ E_g=1 $ as a common energy scale. The momentum cutoff $\delta_k$ is taken to be small enough to guarantee the validity of RWA. We assess this  by solving the full dynamics in the absence of superconductor pairing and comparing with the solutions obtained from the RWA equations in the same regime. For the values of  momentum range $\delta k$ that we use,  the steady-state $k$-space averaged popoulation of the valence band differ between the two solutions by $\mathcal{O}\left(10^{-2}\right)$.

\section{Results}\label{sec:results}
\subsection{Dynamics and Phase diagram}
\begin{figure}[h]
	\centering
	\includegraphics[width=1\linewidth]{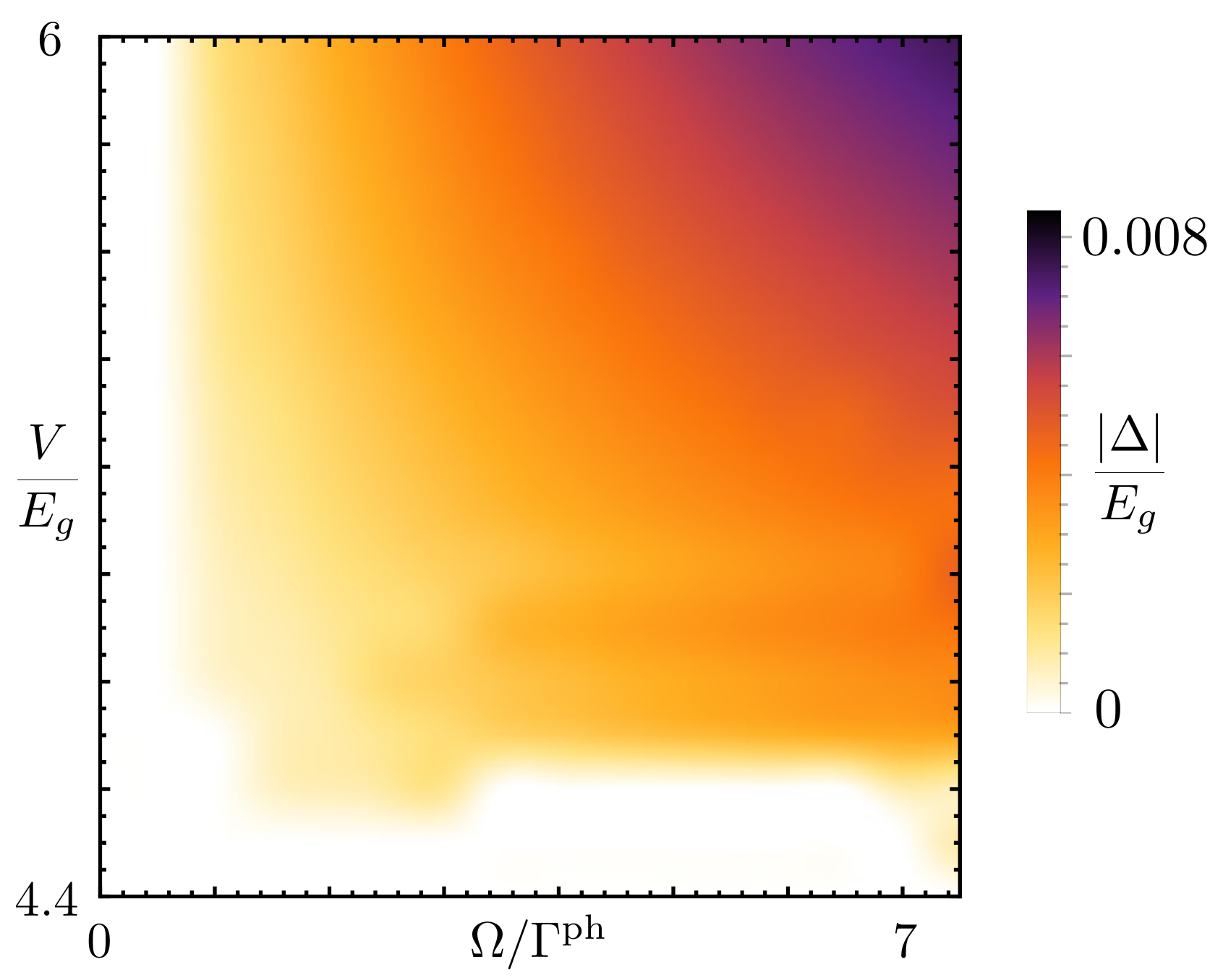}
	\caption{Phase diagram of the system, representing the absolute value of the complex order parameter $ |\Delta| $ as a function of the laser intensity $ \Omega $ and the repulsive interaction strength $ V $. Here, $ \Gamma^{\mathrm{ph}}=0.02E_g $ is the phonon rate, $ \Gamma^{\mathrm{rr}}=\Gamma^{\mathrm{ph}}/10 $ is the radiative recombination rate and $ A_1=7E_g  $, $ A_2=24E_g  $ and we consider a momentum cutoff such that $ |k|\leq\delta_k=0.2 $. }
	\label{fig:PhaseDiagram}
\end{figure}

We solve the dynamics for the relevant observables, i.e. the populations of valence $ n^{11}_k=\braket{c^{1\dag}_k c^1_k} $ and conduction $ n^{22}_k=\braket{c^{2\dag}_k c^2_k} $ bands, and the ordinary $ n^{21}_k=\braket{c^{2\dag}_k c^1_k} $ as well as the anomalous $ s^{21}_k=\braket{c^{2\dag}_k c^{1\dag}_k} $ interband correlations. Their complete time evolution is shown in App.~\ref{app:solution}. For clarity, we focus only on the zero temperature regime ($n_B\tmmathbf{q} =0$), since finite temperature corrections (for $ k_BT\ll E_g $) do not affect qualitatively our results and interpretation. Moreover, we initialize the dynamics with the respective equilibrium state of the system and the baths, where only the valence band is populated. We also have to assume the initial interband anomalous correlations to be non-zero even though very small, in order to avoid the unstable fixed point solution of the dissipative mean field equations $s^{21}_{k} = 0$.\cite{Goldstein_PRB15} Our main result is that, in a rather generic parameter range, the anomalous interband correlator $ s^{21}_k$ reaches a non-zero steady state for any non-zero, but arbitrarily small, initial value. Consequently, a finite interband pairing $\Delta$ can develop in the system. A phase diagram is shown in Fig.~\ref{fig:PhaseDiagram}, where the steady state value of the order parameter is plotted as a function of laser intensity and interaction strength for fixed values of the dissipation rates. Thresholds in both electron-electron interactions and laser intensity are present. Moreover, in the parameter range we could access, both stronger interactions and laser intensity generally imply a larger induced superconducting pairing.\\ It is worth to notice the different roles played by the phononic and the photonic baths. While the phononic bath is crucial in establishing the superconducting steady-state, the radiative recombination does not qualitatively influence the phase diagram, as long as $ {\Gamma^{\mathrm{rr}}=\pi|w|^2<\Omega}. $  Increasing the phononic rate $ \Gamma^{\mathrm{ph}} $, however, qualitatively modifies the value of the superconducting order parameter $ \Delta $ achieved in the stationary state. It moves in fact, the threshold on the interaction strength to larger values. If $ \Gamma^{\mathrm{ph}} $ is raised up even further, superconducting correlations will eventually be washed out. The dependence of the phase diagram on the other parameters is indeed weaker.
\\ In order to better understand the physics it is worth to notice that  the time evolution of the anomalous interband correlator $ s^{21}_k$, and hence the superconducting pairing, is strongly dependent on the quantity $ {\tilde n_k=n^{22}_k+n^{11}_{-k}-1} $ \cite{Goldstein_PRB15,Hart_arxiv18,Note_Supp}. Whenever $\tilde{n}_k$ is close to zero, $s_{k}^{21}$ behaves accordingly\cite{Note_Supp}.  Heuristically, the condition  $ \tilde n_k=0$ implies zero probability of forming a Cooper pair of electrons (or holes) between momenta $k$ and $-k$. However  $ \tilde n_k\neq 0$ is only a necessary condition for obtaining $\Delta \neq 0$.  The condition  ${\tilde{n}_k \neq 0}$ cannot be realized by means of photonic dissipation alone, since there is no significant momentum transfer in the electron sector as a result of such processes. On the other hand, crucially, phonon scattering tends to place electrons at the bottom of the bands, due to the fact that the bands have the same concavity at the resonance point. Hence, it is phonon scattering that generates the condition  $ \tilde n_k\neq 0$ necessary for the development of the superconducting correlations. 
\begin{figure}[h]
	\centering
	\includegraphics[width=1\linewidth]{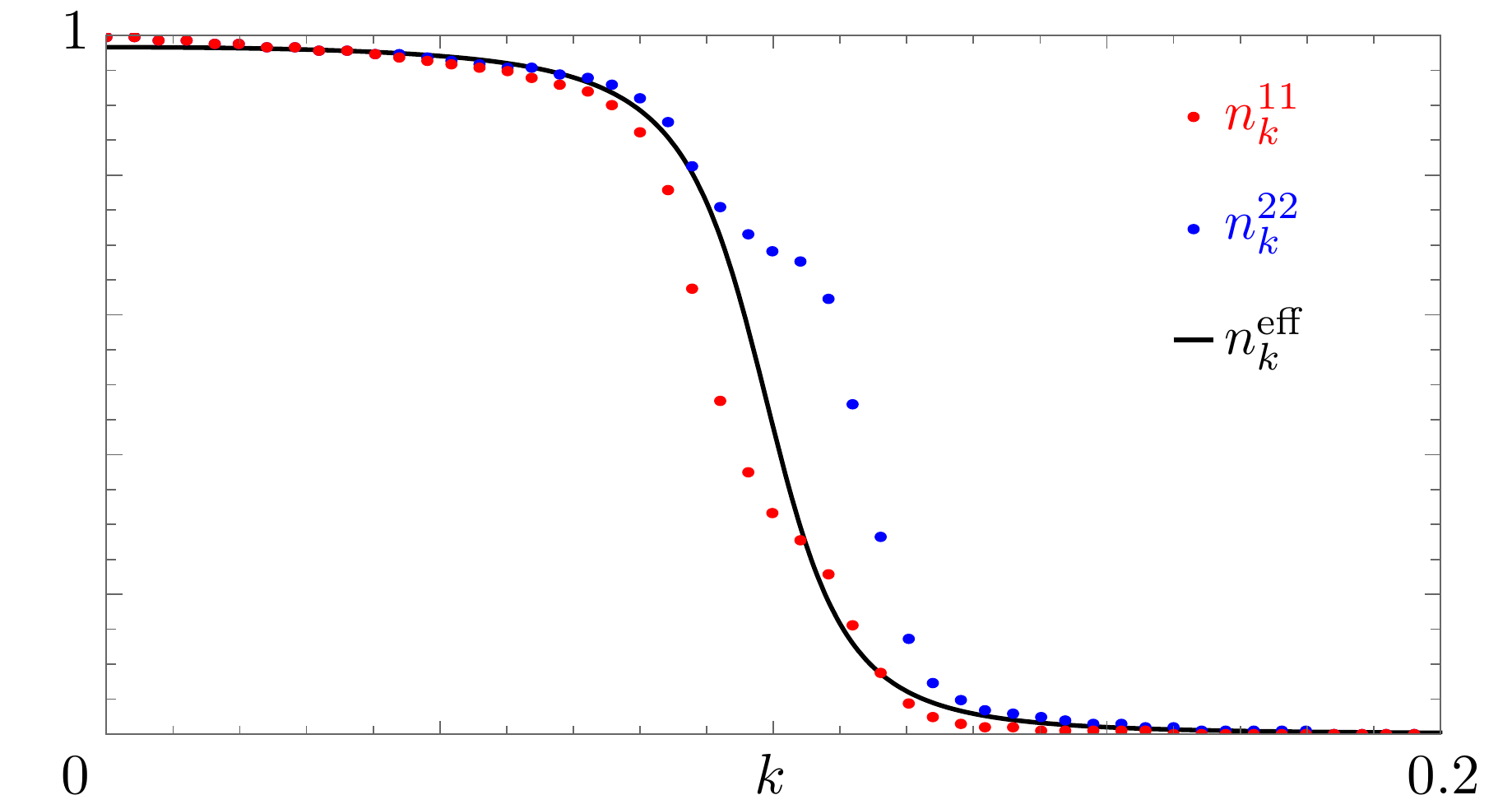}
	\caption{Steady state population of valence (blue dots) and conduction (red dots) bands as a function of the momentum $ k $. We choose $ \Omega/\Gamma^{\mathrm{ph}}= 25 $, where $ \Gamma^{\mathrm{ph}}$ is the phonon rate, $ \Gamma^{\mathrm{rr}}=\Gamma^{\mathrm{ph}}/10 $ is the radiative recombination rate, $ \Omega=0.25E_g  $, $ A_1=7E_g  $, $ A_2=24E_g  $, $V=5 E_g$. We also consider a momentum cutoff  such that $ |k|\leq\delta_k=0.2 $ . The black solid line represents the population of the two bands in the effective system~\cite{Moreo_PRB09} where attractive interaction of strenght $ U=-0.05 $ are present (See Sec.~\ref{sec:effMec} ).}
	\label{fig:nkBCS}
\end{figure}
 The behavior of the populations is shown in Fig.~\ref{fig:nkBCS}. We find in fact $\tilde{n}_k\simeq 1 $ for the momenta around the minima of the quadratic bands while $ \tilde{n}_k\simeq-1 $ on the edge of the region where the rotating wave applies. Note that the transition between these two regions is smooth: In the next section we comment how this feature becomes crucial in the explanation of Cooper pair creation. Qualitatively speaking, the smoothness signals the superconducting interband pairing.
 
\begin{figure}[h]
	\centering
	\includegraphics[width=1\linewidth]{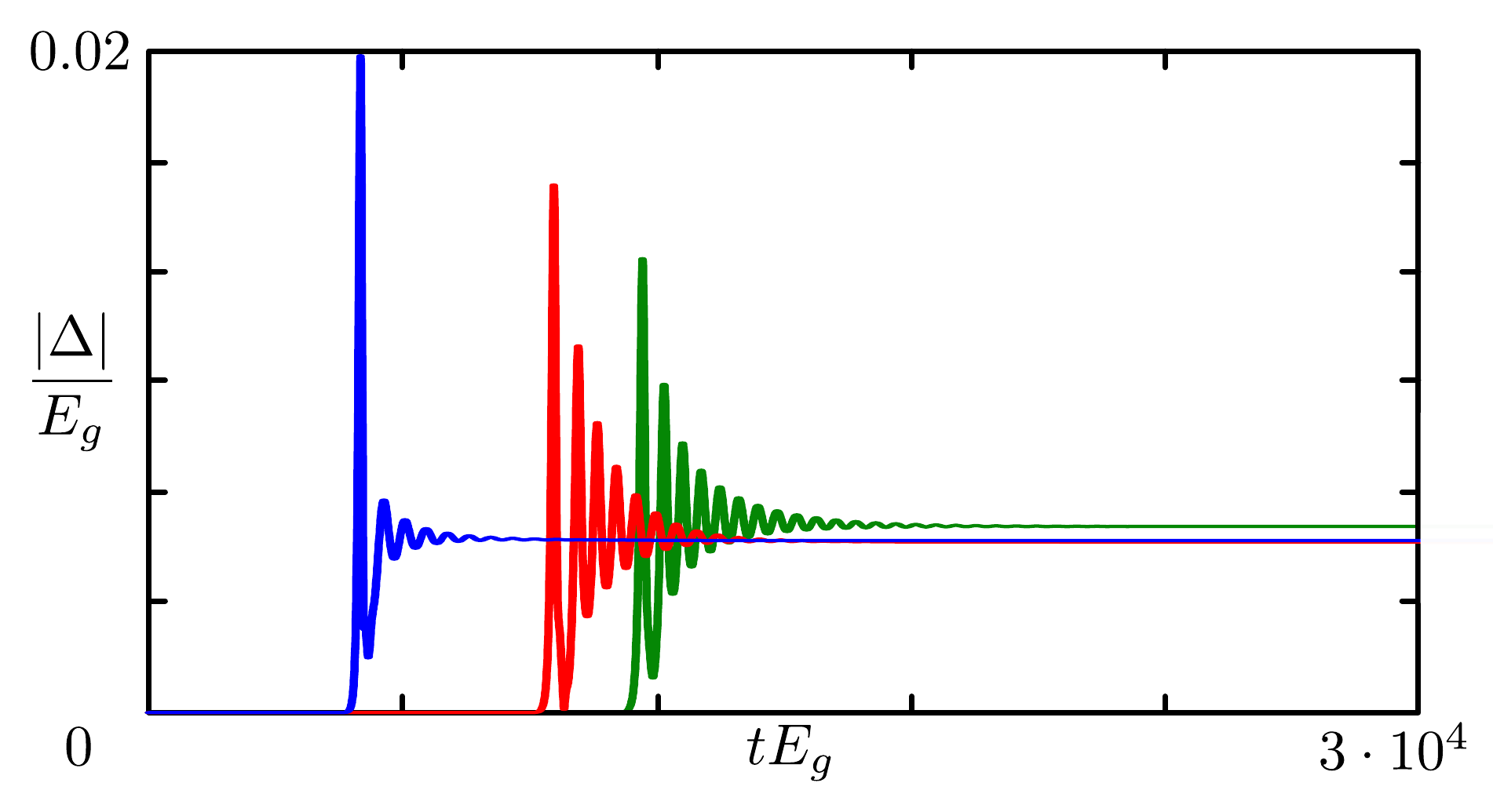}
	\caption{Time evolution of the absolute value of the complex order parameter $ |\Delta| $, for different values of the radiative recombination rate: $ \Gamma^{\mathrm{rr}}=\Gamma^{\mathrm{ph}}/20 $ (green line), $ \Gamma^{\mathrm{rr}}=\Gamma^{\mathrm{ph}}/10 $ (red line), and $ \Gamma^{\mathrm{rr}}=\Gamma^{\mathrm{ph}}/2 $ (blue line). We choose $ \Omega/\Gamma^{\mathrm{ph}}= 25 $, where $ \Gamma^{\mathrm{ph}}$ is the phonon rate, and $ \Omega=0.25E_g  $, $ A_1=7E_g  $, $ A_2=24E_g  $, $ V=4E_g  $. The momentum cutoff is $ |k|\leq \delta_k=0.2 $.} 
	\label{fig:Delta}
\end{figure}

In order to understand the role played by the radiative recombination, we show in Fig.~\ref{fig:Delta} the real time evolution of the complex order parameter $ |\Delta| $. The larger the photonic relaxation rate, the shorter the time needed to reach the steady state. This behavior can be understood by noticing that photonic relaxation adds a qualitatively new (interband) dissipation channel. 
 So, while the steady state populations are not substantially affected by the photonic relaxation, the time needed to establish them changes (see Fig.~\ref{fig:Delta}). 
\subsection{Effective pairing mechanism}\label{sec:effMec}
As shown in Fig.~\ref{fig:nkBCS}, in the superconducting steady state the bands are qualitatively empty at the edges of the considered momentum region and filled in the center, around the minima. We can mimic this situation by employing an equilibrium system, with two bands with positive concavity and coincident minima. Indeed, the action of the laser is to effectively shift the lower band up to the upper one, balancing the gap energy difference between different band electrons. An attractive interaction of the form $ {-U(n^{11}_k-1/2)(n^{22}_{-k}-1/2)} $, with $ U>0 $ favours the situation where, if the state $ k $ in band 1 is occupied (empty), the corresponding state $ -k $ in band 2 is occupied (empty). The chemical potential then determines whether the ground state is given by the couple of empty or filled states. If we apply the mean field approximation to the above-mentioned attractive interaction term, we obtain exactly the model studied in Ref.~\cite{Moreo_PRB09}, which reads:
\begin{equation}\label{eq:bcs}
H_{eff}=\sum_{k,\alpha} \epsilon_{\alpha}(k) c^{\alpha\dag}_{k}c_{k}^\alpha-\tilde\Delta\sum_{k,\alpha\neq\beta}\left(c^{\alpha\dag}_{k}c^{\beta\dag}_{-k}+h.c.\right)
\end{equation}
where $ \alpha,\beta=1,2 $ labels the two bands, $ {\epsilon_{\alpha}(k)=A_\alpha k^2-C} $ and $ \tilde{\Delta}=U\Delta$. We fix $ C\simeq0.2 E_g $ in such a way that the Fermi energy of the effective model at $U=0$ is compatible with the occupation numbers characterizing the steady state reached by the driven-dissipative model when $V=0$. Using a Bogoliubov-de Gennes representation, we are able to diagonalize $H_{eff}$  and evaluate the expectation value of the population. The two populations are approximately equal for the values of $\tilde\Delta$ considered here, because we are in the regime of \textit{strong pairing}~\cite{Moreo_PRB09}. Therefore we consider only one of the two and call it $n_k^{\mathrm{eff}}$  (see black solid line in Fig.~\ref{fig:nkBCS}). We observe a very similar qualitative behavior between $n_k^{\mathrm{eff}}$   and the nonequilibrium steady state populations of our original driven-dissipative model. Therefore, the steady state can be effectively seen as an equilibrium two band system where the laser light shifts the valence band up to the conduction band and an attractive interaction creates Cooper pairs. This simple effective model is intended to clarify the reason why the condition $\tilde{n}_k\neq 0$ is essential but not sufficient for the superconducting pairing. In fact, a simpler model with two bands and two different chemical potentials would lead to $ \tilde{n_k}\neq0 $, but would not imply coupling between the bands and consequently the smearing of the populations as well as superconductivity.
A full qualitative interpretation of the rather involved results can hence be given: The laser excites electrons from the valence to the conduction band, at any $k$ in the resonance region. The phonons let the electrons in each band get as close as possible to the bottom of the bands (the phononic bath is hence crucial). Interactions between the two bands drive the system into a correlated state. A possible mean field effective model describing the steady state reached is provided by the simple Bogoliubov-de Gennes Hamiltonian for interband superconductors, given in Eq.~\eqref{eq:bcs}.
\section{Conclusions}
We have shown that a non-equilibrium steady state characterized by a finite superconducting order parameter related to an anomalous interband pairing can be achieved in a semiconductor by coupling the bands with  laser light. Specifically, we have considered quadratic bands with the same sign of the effective mass. We have shown that an acoustic phononic bath, which is responsible of intraband transitions only, can induce an electronic distribution in the valence and conduction bands which favours the development of such an unusual pairing. Remarkably, this picture is not destroyed even if an interband relaxation with the same order of magnitude as the intraband one is switched on. Indeed, interband relaxation can speed up the formation of the superconducting order parameter. Furthermore, the stationary state with $ |\Delta|\neq 0 $ can be achieved only for sufficiently large repulsive density-density electronic interaction ($ V\gtrsim 2E_g  $) and for an initial configuration where $ |\Delta| $ can be vanishing compatibly with the numerical precision. The Rabi frequency $ \Omega $, on the other hand, has to be strong enough ($ {\Omega > \Gamma^{\mathrm{ph}}} $), since the laser has to be able to drive a sufficient number of electrons in the conduction band. Only in this case, the two populations can be forced to satisfy the favourable condition $ \tilde{n_k}\neq0 $.  The result is stable against changing the velocity $ v $ of the acoustic phonons, the amplitude of the bands and the $k$-space discretization step. Moreover, the phonon coupling $ \Gamma^{\mathrm{ph}} $, as well as the radiative recombination strength $ \Gamma^{\mathrm{rr}} $, have been chosen to represent a realistic scenario: Phonon transitions in prototypical semiconductors such as Silicon or Germanium have tipically a lifetime $\tau_{\mathrm{ph}}=0.1 ps$, which correspond to a rate $\Gamma_{\mathrm{ph}}\sim 0.01 E_g$ \cite{Tandon_JAP15}, while  radiative recombination rates are strongly dependent on the electron concentration~\cite{coldren2012diode}. However, their action does not qualitatively affect the steady state reached by the system.  Finally, we argue that our results corroborate the original proposal of Goldstein et al.\cite{Goldstein_PRB15}, by putting it in a more realistic framework. This fact, together with the interpretation of the non-equilibrium pairing based on equilibrium multiband BCS theory, should facilitate the experimental detection of this novel route to light-induced insulator/superconductor transition.

\acknowledgments

This work was supported by the DFG (SPP1666, SFB1170, and the W\"urzburg-Dresden Cluster of Excellence ct.qmat, EXC2147).\\
*S. P. and L. P. contributed equally to this work. L.P. thanks A. Nava for discussions.

\newpage
\appendix

\section{Dissipative time evolution equations for the observables}
In this section, we outline the derivation of the equations of motion for the relevant observables of the system. Since the coherent part of Eq.~\eqref{eq:dto} is identical to the one in Ref.~\cite{Goldstein_PRB15},
we will focus here only on the dissipative term of the Lindblad equation.\\
\subsection{Deriving the equations of motion}\label{app:eom}
We focus on the explicit derivation of the dynamics of $n^{11}_{\tmmathbf{p}}={c^{1
		\dag}_{\tmmathbf{p}} c^1_{\tmmathbf{p}}} $. We start with the phonon dissipation. Exploiting the commutators in Eq.~\eqref{eq:dto} and according to the definition in Eq.~\eqref{eq:defS} one finds
\begin{widetext}	
\begin{equation}
\left[\mathcal{S}^{\alpha}_{\tmmathbf{k}, \tmmathbf{q}}, n^{11}_{\tmmathbf{p}}\right] = \delta_{\alpha, 1} \left[\delta
(\tmmathbf{k}+\tmmathbf{q}-\tmmathbf{p}) c_{\tmmathbf{p}-\tmmathbf{q}}^{1 \dag} c_{\tmmathbf{p}}^1 - \delta (\tmmathbf{k}-\tmmathbf{p})
c_{\tmmathbf{p}}^{1 \dag} c_{\tmmathbf{p}+\tmmathbf{q}}^1\right]\,,
\end{equation}
\begin{equation}
\left[n^{11}_{\tmmathbf{p}},\mathcal{S}^{\alpha\dag}_{\tmmathbf{k}, \tmmathbf{q}}\right] = \delta_{\alpha, 1} \left[\delta
(\tmmathbf{k}+\tmmathbf{q}-\tmmathbf{p}) c_{\tmmathbf{p}}^{1 \dag} c_{\tmmathbf{p}-\tmmathbf{q}}^1 - \delta (\tmmathbf{k}-\tmmathbf{p})
c_{\tmmathbf{p}+\tmmathbf{q}}^{1 \dag} c_{\tmmathbf{p}}^1\right]\,.
\end{equation}
\end{widetext}
We are now in a position to evaluate the quantum averages in Eq.~\eqref{eq:dto}. For simplicity we omit the explicit dependence of the bosonic populations $n_B$ on $\tmmathbf{q}$. The basic elements of the terms with $1+n_B$ as a prefactor are
\begin{widetext}
\begin{equation}
\langle c_{\tmmathbf{k}'}^{\beta \dag} c_{\tmmathbf{k}'+\tmmathbf{q}}^{\beta} c_{\tmmathbf{p}}^{1 \dag}c_{\tmmathbf{p}-\tmmathbf{q}}^{1}\rangle =\delta_{\beta, 2}\delta(\tmmathbf{k}'+\tmmathbf{p})\braket{s^{21}_{-\tmmathbf{p}}}\braket{s^{21\dag}_{-\tmmathbf{p}+\tmmathbf{q}}}+\delta(\tmmathbf{k}'-\tmmathbf{p}+\tmmathbf{q})\left[\delta_{\beta, 1}\braket{n^{11}_{\tmmathbf{p}-\tmmathbf{q}}}\braket{1-n^{11}_{\tmmathbf{p}}}-\delta_{\beta, 2}\braket{n^{21}_{\tmmathbf{p}-\tmmathbf{q}}}\braket{n^{12}_{\tmmathbf{p}}}\right] \;,
\end{equation}
\begin{equation}
\langle c_{\tmmathbf{k}'}^{\beta \dag} c_{\tmmathbf{k}'+\tmmathbf{q}}^{\beta} c_{\tmmathbf{p}+\tmmathbf{q}}^{1 \dag}c_{\tmmathbf{p}}^{1}\rangle =\delta_{\beta, 2}\delta(\tmmathbf{k}'+\tmmathbf{p}+\tmmathbf{q})\braket{s^{21}_{-\tmmathbf{p}-\tmmathbf{q}}}\braket{s^{21\dag}_{-\tmmathbf{p}}}+\delta(\tmmathbf{k}'-\tmmathbf{p})\left[\delta_{\beta, 1}\braket{n^{11}_{\tmmathbf{p}}}\braket{1-n^{11}_{\tmmathbf{p}+\tmmathbf{q}}}-\delta_{\beta, 2}\braket{n^{21}_{\tmmathbf{p}}}\braket{n^{12}_{\tmmathbf{p}+\tmmathbf{q}}}\right] \;,
\end{equation}
\end{widetext}
while the ones with $ n_B $ as a prefactor are
\begin{widetext}
\begin{equation}
\langle c_{\tmmathbf{k}'+\tmmathbf{q}}^{\beta \dag} c_{\tmmathbf{k}'}^{\beta} c_{\tmmathbf{p}-\tmmathbf{q}}^{1 \dag}c_{\tmmathbf{p}}^{1}\rangle= \delta_{\beta, 2}\delta(\tmmathbf{k}'+\tmmathbf{p})\braket{s^{21}_{-\tmmathbf{p}+\tmmathbf{q}}}\braket{s^{21\dag}_{-\tmmathbf{p}}}+\delta(\tmmathbf{k}'-\tmmathbf{p}+\tmmathbf{q})\left[\delta_{\beta, 1}\braket{n^{11}_{\tmmathbf{p}}}\braket{1-n^{11}_{\tmmathbf{p}-\tmmathbf{q}}}-\delta_{\beta, 2}\braket{n^{21}_{\tmmathbf{p}}}\braket{n^{12}_{\tmmathbf{p}-\tmmathbf{q}}}\right] \;,
\end{equation}
\begin{equation}
\langle c_{\tmmathbf{k}'+\tmmathbf{q}}^{\beta \dag} c_{\tmmathbf{k}'}^{\beta} c_{\tmmathbf{p}}^{1 \dag}c_{\tmmathbf{p}+\tmmathbf{q}}^{1}\rangle =\delta_{\beta, 2}\delta(\tmmathbf{k}'+\tmmathbf{p}+\tmmathbf{q})\braket{s^{21}_{-\tmmathbf{p}}}\braket{s^{21\dag}_{-\tmmathbf{p}-\tmmathbf{q}}}+\delta(\tmmathbf{k}'-\tmmathbf{p})\left[\delta_{\beta, 1}\braket{n^{11}_{\tmmathbf{p}+\tmmathbf{q}}}\braket{1-n^{11}_{\tmmathbf{p}}}-\delta_{\beta, 2}\braket{n^{21}_{\tmmathbf{p}+\tmmathbf{q}}}\braket{n^{12}_{\tmmathbf{p}}}\right] \;,
\end{equation}
\end{widetext}
Collecting all terms we finally obtain for the phononic dissipation
\begin{widetext}
\begin{equation}
\begin{aligned}
\frac{d}{d t} \langle n^{11}_{\tmmathbf{p}} \rangle^{\tmop{ph}} =  2\sum_{
	\tmmathbf{q}} & \left\{ [1 + n_B] \left[\Gamma_{\tmmathbf{p}-\tmmathbf{q},\tmmathbf{q}}^1 \braket{n^{11}_{\tmmathbf{p}-\tmmathbf{q}}}\braket{1-n^{11}_{\tmmathbf{p}}}-\Gamma_{\tmmathbf{p}-\tmmathbf{q},\tmmathbf{q}}^2 \mathrm{Re}\left\{ \braket{n^{21}_{\tmmathbf{p}-\tmmathbf{q}}}\braket{n^{12}_{\tmmathbf{p}}}\right\}-\Gamma_{-\tmmathbf{p},\tmmathbf{q}}^2 \mathrm{Re}\left\{ \braket{s^{21}_{-\tmmathbf{p}}}\braket{s^{21\dag}_{-\tmmathbf{p}+\tmmathbf{q}}}\right\}\right]\right.+ \\&
-[1 + n_B] \left[\Gamma_{\tmmathbf{p},\tmmathbf{q}}^1 \braket{n^{11}_{\tmmathbf{p}}}\braket{1-n^{11}_{\tmmathbf{p}+\tmmathbf{q}}}-\Gamma_{\tmmathbf{p},\tmmathbf{q}}^2 \mathrm{Re}\left\{ \braket{n^{21}_{\tmmathbf{p}}}\braket{n^{12}_{\tmmathbf{p}+\tmmathbf{q}}}\right\}-\Gamma_{-\tmmathbf{p}-\tmmathbf{q},\tmmathbf{q}}^2 \mathrm{Re}\left\{ \braket{s^{21}_{-\tmmathbf{p}-\tmmathbf{q}}}\braket{s^{21\dag}_{-\tmmathbf{p}}}\right\}\right] + \\&
+n_B \left[\Gamma_{\tmmathbf{p}-\tmmathbf{q},\tmmathbf{q}}^1 \braket{n^{11}_{\tmmathbf{p}}}\braket{1-n^{11}_{\tmmathbf{p}-\tmmathbf{q}}}-\Gamma_{\tmmathbf{p}-\tmmathbf{q},\tmmathbf{q}}^2 \mathrm{Re}\left\{ \braket{n^{21}_{\tmmathbf{p}}}\braket{n^{12}_{\tmmathbf{p}-\tmmathbf{q}}}\right\}-\Gamma_{-\tmmathbf{p},\tmmathbf{q}}^2 \mathrm{Re}\left\{ \braket{s^{21}_{-\tmmathbf{p}+\tmmathbf{q}}}\braket{s^{21\dag}_{-\tmmathbf{p}}}\right\}\right]+  \\&
-\left. n_B \left[\Gamma_{\tmmathbf{p},\tmmathbf{q}}^1 \braket{n^{11}_{\tmmathbf{p}+\tmmathbf{q}}}\braket{1-n^{11}_{\tmmathbf{p}}}-\Gamma_{\tmmathbf{p},\tmmathbf{q}}^2 \mathrm{Re}\left\{ \braket{n^{21}_{\tmmathbf{p}+\tmmathbf{q}}}\braket{n^{12}_{\tmmathbf{p}}}\right\}-\Gamma_{-\tmmathbf{p}-\tmmathbf{q},\tmmathbf{q}}^2 \mathrm{Re}\left\{ \braket{s^{21}_{-\tmmathbf{p}}}\braket{s^{21\dag}_{-\tmmathbf{p}-\tmmathbf{q}}}\right\}\right]\right\} .
\end{aligned}
\end{equation}
\end{widetext}
The dissipative part due to the radiative recombination follows analogously and reads
\begin{equation}
\begin{aligned}
\frac{d}{d t} \langle n^{11}_{\tmmathbf{p}} \rangle^{\tmop{rr}} = \Gamma^{\tmop{rr}} \langle n^{22}_{\tmmathbf{p}} \rangle \langle 1-n^{11}_{\tmmathbf{p}} \rangle ,
\end{aligned}
\end{equation}
see Eq.~\eqref{eq:Grr}.\\

\noindent Therefore, the full equation of motion for $n^{11}_{\tmmathbf{p}}$ reads
\begin{widetext}
\begin{equation}\label{eq:n11}
\begin{aligned}
\frac{d}{d t} \langle n^{11}_{\tmmathbf{p}} \rangle =  &-\Omega  \mathrm{ Im}\left\{\braket{n^{21}_{\tmmathbf{p}}}\right\}-2 \mathrm{ Im}\left\{\Delta \braket{s^{21}_{\tmmathbf{p}}}\right\} + 
\Gamma^{\tmop{rr}} \langle n^{22}_{\tmmathbf{p}} \rangle \langle 1-n^{11}_{\tmmathbf{p}} \rangle + & \\ &+2\sum_{
	\tmmathbf{q}}  \left\{  \left[\Gamma_{\tmmathbf{p}-\tmmathbf{q},\tmmathbf{q}}^1 \braket{n^{11}_{\tmmathbf{p}-\tmmathbf{q}}}\braket{1-n^{11}_{\tmmathbf{p}}}-\Gamma_{\tmmathbf{p}-\tmmathbf{q},\tmmathbf{q}}^2 \mathrm{Re}\left\{ \braket{n^{21}_{\tmmathbf{p}-\tmmathbf{q}}}\braket{n^{12}_{\tmmathbf{p}}}\right\}-\Gamma_{-\tmmathbf{p},\tmmathbf{q}}^2 \mathrm{Re}\left\{ \braket{s^{21}_{-\tmmathbf{p}}}\braket{s^{21\dag}_{-\tmmathbf{p}+\tmmathbf{q}}}\right\}\right]\right.+ & \\
&\qquad\qquad-\left. \left[\Gamma_{\tmmathbf{p},\tmmathbf{q}}^1 \braket{n^{11}_{\tmmathbf{p}}}\braket{1-n^{11}_{\tmmathbf{p}+\tmmathbf{q}}}-\Gamma_{\tmmathbf{p},\tmmathbf{q}}^2 \mathrm{Re}\left\{ \braket{n^{21}_{\tmmathbf{p}}}\braket{n^{12}_{\tmmathbf{p}+\tmmathbf{q}}}\right\}-\Gamma_{-\tmmathbf{p}-\tmmathbf{q},\tmmathbf{q}}^2 \mathrm{Re}\left\{ \braket{s^{21}_{-\tmmathbf{p}-\tmmathbf{q}}}\braket{s^{21\dag}_{-\tmmathbf{p}}}\right\}\right] \right\}\,. &
\end{aligned}
\end{equation}
The equations of motion for $n^{22}_{\tmmathbf{p}}$, $n^{21}_{\tmmathbf{p}}$ and $s^{21}_{\tmmathbf{p}}$ can be derived along the very same lines as above. We quote here the final results
\begin{equation}\label{eq:n22}
\begin{aligned}
\frac{d}{d t} \langle n^{22}_{\tmmathbf{p}} \rangle =& +\Omega  \mathrm{ Im}\left\{\braket{n^{21}_{\tmmathbf{p}}}\right\}-2 \mathrm{ Im}\left\{\Delta \braket{s^{21}_{\tmmathbf{p}}}\right\} 
-\Gamma^{\tmop{rr}} \langle n^{22}_{\tmmathbf{p}} \rangle \langle 1-n^{11}_{\tmmathbf{p}} \rangle +& \\ & +2\sum_{
	\tmmathbf{q}}  \left\{  \left[\Gamma_{\tmmathbf{p}-\tmmathbf{q},\tmmathbf{q}}^2 \braket{n^{22}_{\tmmathbf{p}-\tmmathbf{q}}}\braket{1-n^{22}_{\tmmathbf{p}}}-\Gamma_{\tmmathbf{p}-\tmmathbf{q},\tmmathbf{q}}^1 \mathrm{Re}\left\{ \braket{n^{12}_{\tmmathbf{p}-\tmmathbf{q}}}\braket{n^{21}_{\tmmathbf{p}}}\right\}-\Gamma_{-\tmmathbf{p},\tmmathbf{q}}^1 \mathrm{Re}\left\{ \braket{s^{12}_{-\tmmathbf{p}}}\braket{s^{12\dag}_{-\tmmathbf{p}+\tmmathbf{q}}}\right\}\right]\right.+ & \\
&\qquad\qquad-\left.\left[\Gamma_{\tmmathbf{p},\tmmathbf{q}}^2 \braket{n^{22}_{\tmmathbf{p}}}\braket{1-n^{22}_{\tmmathbf{p}+\tmmathbf{q}}}-\Gamma_{\tmmathbf{p},\tmmathbf{q}}^1 \mathrm{Re}\left\{ \braket{n^{12}_{\tmmathbf{p}}}\braket{n^{21}_{\tmmathbf{p}+\tmmathbf{q}}}\right\}-\Gamma_{-\tmmathbf{p}-\tmmathbf{q},\tmmathbf{q}}^1 \mathrm{Re}\left\{ \braket{s^{12}_{-\tmmathbf{p}-\tmmathbf{q}}}\braket{s^{12\dag}_{-\tmmathbf{p}}}\right\}\right] \right\}\,, 
\end{aligned}
\end{equation}

\begin{equation}\label{eq:n21}
\begin{aligned}
\frac{d}{d t} \langle n^{21}_{\tmmathbf{p}} \rangle =& - \dfrac{i \Omega}{2} \braket{n^{22}_{\tmmathbf{p}}-n^{11}_{\tmmathbf{p}}}+i \epsilon(\tmmathbf{p})\braket{n^{21}_{\tmmathbf{p}}} -\Gamma^{\tmop{rr}} \langle n^{21}_{\tmmathbf{p}} \rangle \langle 1-n^{11}_{\tmmathbf{p}}+n^{22}_{\tmmathbf{p}} \rangle +&\\& + \sum_{
	\tmmathbf{q}}  \left\{  \left[\Gamma_{\tmmathbf{p}-\tmmathbf{q},\tmmathbf{q}}^1 \left(\braket{n^{21}_{\tmmathbf{p}-\tmmathbf{q}}}\braket{1-n^{11}_{\tmmathbf{p}}}-\braket{n^{11}_{\tmmathbf{p}-\tmmathbf{q}}}\braket{n^{21}_{\tmmathbf{p}}}\right)+\Gamma_{\tmmathbf{p}-\tmmathbf{q},\tmmathbf{q}}^2 \left(\braket{n^{21}_{\tmmathbf{p}-\tmmathbf{q}}}\braket{1-n^{22}_{\tmmathbf{p}}}-\braket{n^{22}_{\tmmathbf{p}-\tmmathbf{q}}}\braket{n^{21}_{\tmmathbf{p}}}\right)\right]\right.+ & \\
& \qquad\quad +  \left.\left[\Gamma_{\tmmathbf{p},\tmmathbf{q}}^1 \left(\braket{n^{11}_{\tmmathbf{p}}}\braket{n^{21}_{\tmmathbf{p}+\tmmathbf{q}}}-\braket{n^{21}_{\tmmathbf{p}}}\braket{1-n^{11}_{\tmmathbf{p}+\tmmathbf{q}}}\right)+\Gamma_{\tmmathbf{p},\tmmathbf{q}}^2 \left(\braket{n^{22}_{\tmmathbf{p}}}\braket{n^{21}_{\tmmathbf{p}+\tmmathbf{q}}}-\braket{n^{21}_{\tmmathbf{p}}}\braket{1-n^{22}_{\tmmathbf{p}+\tmmathbf{q}}}\right)\right] \right\}\,,
\end{aligned}
\end{equation}

\begin{equation}\label{eq:s21}
\begin{aligned}
\frac{d}{d t} \langle s^{21}_{\tmmathbf{p}} \rangle = &+i \Delta^* \braket{n^{22}_{\tmmathbf{p}}+n^{11}_{-\tmmathbf{p}}-1}+i E(\tmmathbf{p})\braket{s^{21}_{\tmmathbf{p}}} +&\\ & -\sum_{
	\tmmathbf{q}}  \left\{ \left[\Gamma_{-\tmmathbf{p},\tmmathbf{q}}^1 \left(\braket{s^{21}_{\tmmathbf{p}}}\braket{1-n^{11}_{-\tmmathbf{p}+\tmmathbf{q}}}+\braket{n^{11}_{-\tmmathbf{p}}}\braket{s^{21}_{\tmmathbf{p}-\tmmathbf{q}}}\right)+\Gamma_{\tmmathbf{p}-\tmmathbf{q},\tmmathbf{q}}^2 \left(\braket{s^{21}_{\tmmathbf{p}-\tmmathbf{q}}}\braket{1-n^{22}_{\tmmathbf{p}}}+\braket{n^{22}_{\tmmathbf{p}-\tmmathbf{q}}}\braket{s^{21}_{\tmmathbf{p}}}\right)\right]\right.+ & \\
&\qquad\quad\left.+ \left[\Gamma_{\tmmathbf{p},\tmmathbf{q}}^2 \left(\braket{n^{22}_{\tmmathbf{p}}}\braket{s^{21}_{\tmmathbf{p}+\tmmathbf{q}}}+\braket{s^{21}_{\tmmathbf{p}}}\braket{1-n^{22}_{\tmmathbf{p}+\tmmathbf{q}}}\right)+\Gamma_{-\tmmathbf{p}-\tmmathbf{q},\tmmathbf{q}}^1 \left(\braket{n^{11}_{-\tmmathbf{p}-\tmmathbf{q}}}\braket{s^{21}_{\tmmathbf{p}}}+\braket{s^{21}_{\tmmathbf{p}+\tmmathbf{q}}}\braket{1-n^{11}_{-\tmmathbf{p}}}\right)\right]\right\}\,.
\end{aligned}
\end{equation}
\end{widetext}
where $ \epsilon(\tmmathbf{p})=E_2(\tmmathbf{p})-E_1(\tmmathbf{p})-\nu $, $ E(\tmmathbf{p})=E_2(\tmmathbf{p})+E_1(\tmmathbf{p}) $ and $\Gamma_{\tmmathbf{k}',\tmmathbf{q}}^{\alpha\beta}$ is defined in Eq.~\eqref{eq:gammaph}. Note that the equation for $s^{21}_{\tmmathbf{p}}$ does not depend on $\Gamma^{\tmop{rr}}$, which means that radiative recombination does not affect superconducting correlations.

\section{Solution for quadratic bands at $ T=0 $}\label{app:solution}
Assuming spherical symmetry, so that the $ d $-dimensional model can be reduced to an equivalent one-dimensional model -- see main text, we model the semiconductor bands as
\begin{equation}
E_1(p)=A_{1} p^2-E_g/2 \qquad E_2(p)=A_{2} p^2+E_g/2
\end{equation}
where $ E_g $ represents the gap amplitude in $ p=0 $. Since we are considering the zero temperature case phonons can only be emitted: the only possible transitions are the ones where the electron acquires a momentum $\pm q $ and loses an amount of energy equal to $ v|q| $. According to this picture, if the band concavity is positive then the phonon transitions tend to move the electronic population to the center of the band, while, in the opposite case, the electrons tend to move to the edge of the momentum region considered. \\\\
The   equations of motion displayed in App.~\ref{app:eom} can be solved numerically. In contrast to standard mean-field calculations, there is no need to solve them self-consistently together with an equation for the order parameter.  Indeed the "self-consistency" is taken into account by the non-linearity of our master equation: $\Delta$ is itself a time-dependent quantity. Moreover, we consider the momentum $ p $ as a discrete variable and we convert the energy conservation Dirac deltas in Eq.~\eqref{eq:gammaph} into the momentum representation. As an example we quote here one possible term
\begin{widetext}
\begin{equation}
\begin{split}
\delta \left[E_{\beta}
(p-q)-E_\beta(p)-v|q|\right] =& \dfrac{1}{|2A_\beta p+v|}\delta \left[q-2p- \dfrac{v}{A_\beta}\right]\Theta\left(p+\dfrac{v}{2A_\beta}\right)+\\&\dfrac{1}{|2A_\beta p-v|}\delta \left[q-2p+ \dfrac{v}{A_\beta}\right]\Theta\left(-p+\dfrac{v}{2A_\beta}\right)\,.
\end{split}
\end{equation}
\end{widetext}
When we convert the Dirac deltas from energy to momentum, we obtain the full $k$ dependence of the scattering rate $ \Gamma^{\mathrm{ph}} $, due to both the density of states and the electron-phonon coupling strength $t_{k,q}$. Note that every electronic state $ p $ is influenced by dissipation and phonons induce a constant momentum exchange given by their linear dispersion relation.\\
\noindent In order to mimic realistic low-temperature scenarios, occurring for instance in prototypical semiconductors like Silicon and Germanium~\cite{EPrates}, we set the coupling constants $ t_{k,q} $ such that the corresponding scattering rate $ \Gamma^{\mathrm{ph}} $ is approximately constant over the momentum window close to the resonance. 
\\A problem which arises in this picture is the implementation of the particle number conservation in the system, which has to hold in the $ \Delta=0 $ regime. To fulfil this condition, we ensure that our equations satisfy the principle of detailed balance. 
Furthermore, since we are considering a finite region in the momentum space near the laser resonance, where the rotating wave approximation works, states near the two edges of this box can only lose(receive) particles if the concavity of the band is positive(negative). \\
Note that all the energies are measured with respect to the gap amplitude $ E_g $ at $ k=0 $. We have performed all calculations with a standard 4-th order Runge-Kutta method with adaptive time-step.
\begin{figure}
	\centering
	\includegraphics[width=1\linewidth]{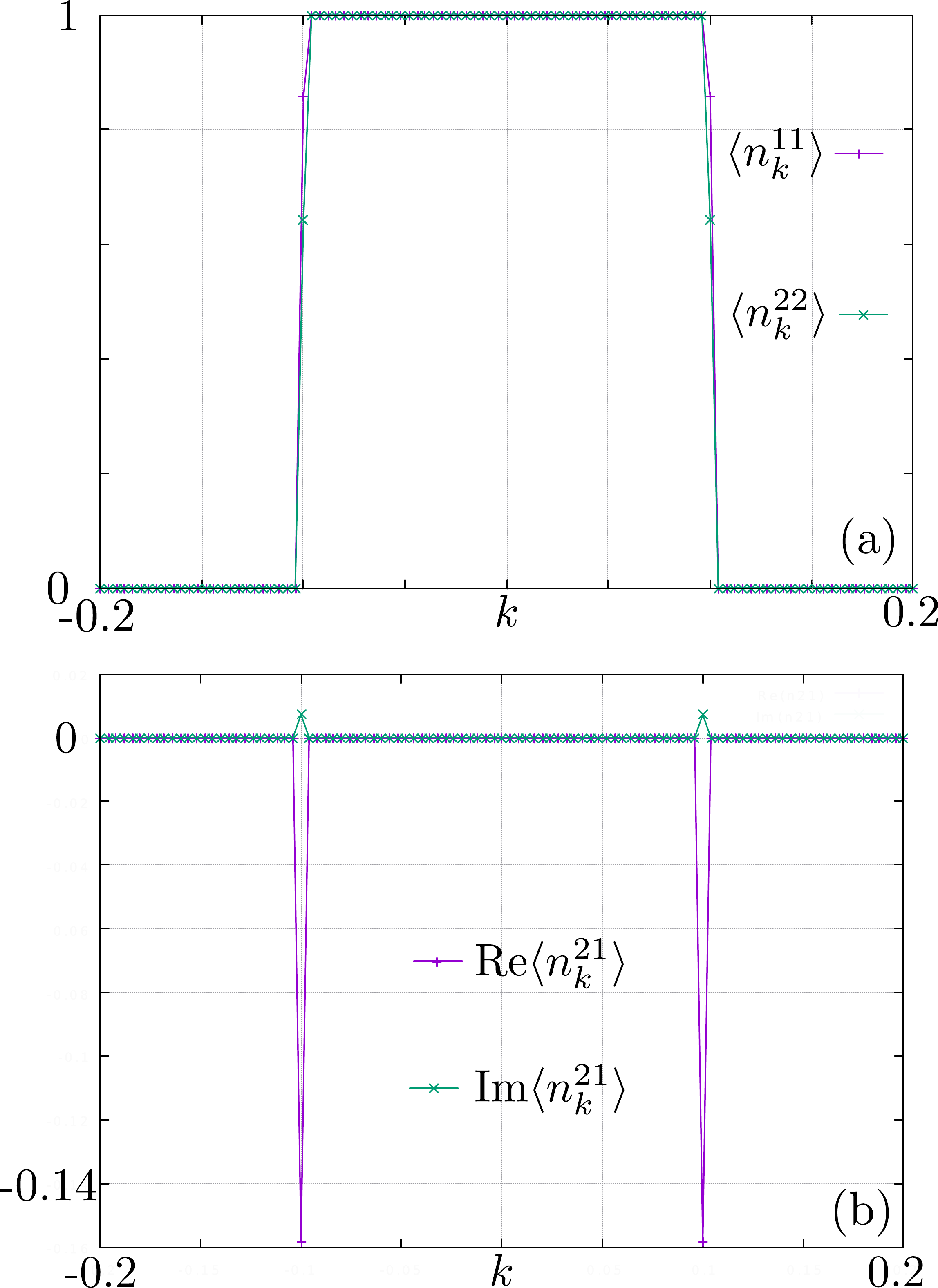}
	\caption{a) Occupation number of valence $ \braket{n^{11}_k} $ and conduction $ \braket{n^{22}_k} $ bands in the stationary state. b) Real and imaginary part of the interband correlation $ \braket{n^{21}_k} $ in the stationary state. Parameters: $ A_1=8E_g $, $ A_2=23E_g $, $ \Gamma^{\tmop{ph}}=0.01E_g $, $ \Gamma^{\tmop{rr}}=\Gamma^{\tmop{ph}} $, and $ \Omega=0.25E_g $. Here, $V=0$ and the momentum cutoff is $ |k|\leq\delta_k=0.2 $. }
	\label{fig:nlaser}
\end{figure}

\subsection{Numerical solution without superconductivity}
In this section, we show the numerical solution of the system in presence of dissipation and electric field. The latter is tuned at the resonant point  $ k=0 $, where both bands have their minimum. In particular, this condition amounts to fixing the laser frequency to $\nu=E_g=1  $. The amplitude of the laser is set to $ \Omega = 0.25E_g $, and the initial conditions coincide with the ground state of the system:  $ \braket{n^{11}_k(t=0)}=1 $, $ \braket{n^{22}_k(t=0)}=0 $, and $ \braket{n^{21}_k(t=0)}=0 $. \\
The resulting stationary state achieved starting from these initial conditions is shown in Figs.~\ref{fig:nlaser} (a) and (b), in the case of  $ \Gamma^{\tmop{rr}}=\Gamma^{\tmop{ph}} $.
\\The stationary state is characterized by a very similarly populated conduction and valence bands, e.g. with electrons occupying the same region around $k=0$ in both of them, On the other hand, the interband correlation vanishes everywhere except at the border between the occupied and empty regions. Analogous results are obtained by $ \Gamma^{\tmop{rr}}=\Gamma^{\tmop{ph}}/10 $. 

\subsection{Complete numerical solution}
\begin{figure}
	\centering
	\includegraphics[width=1\linewidth]{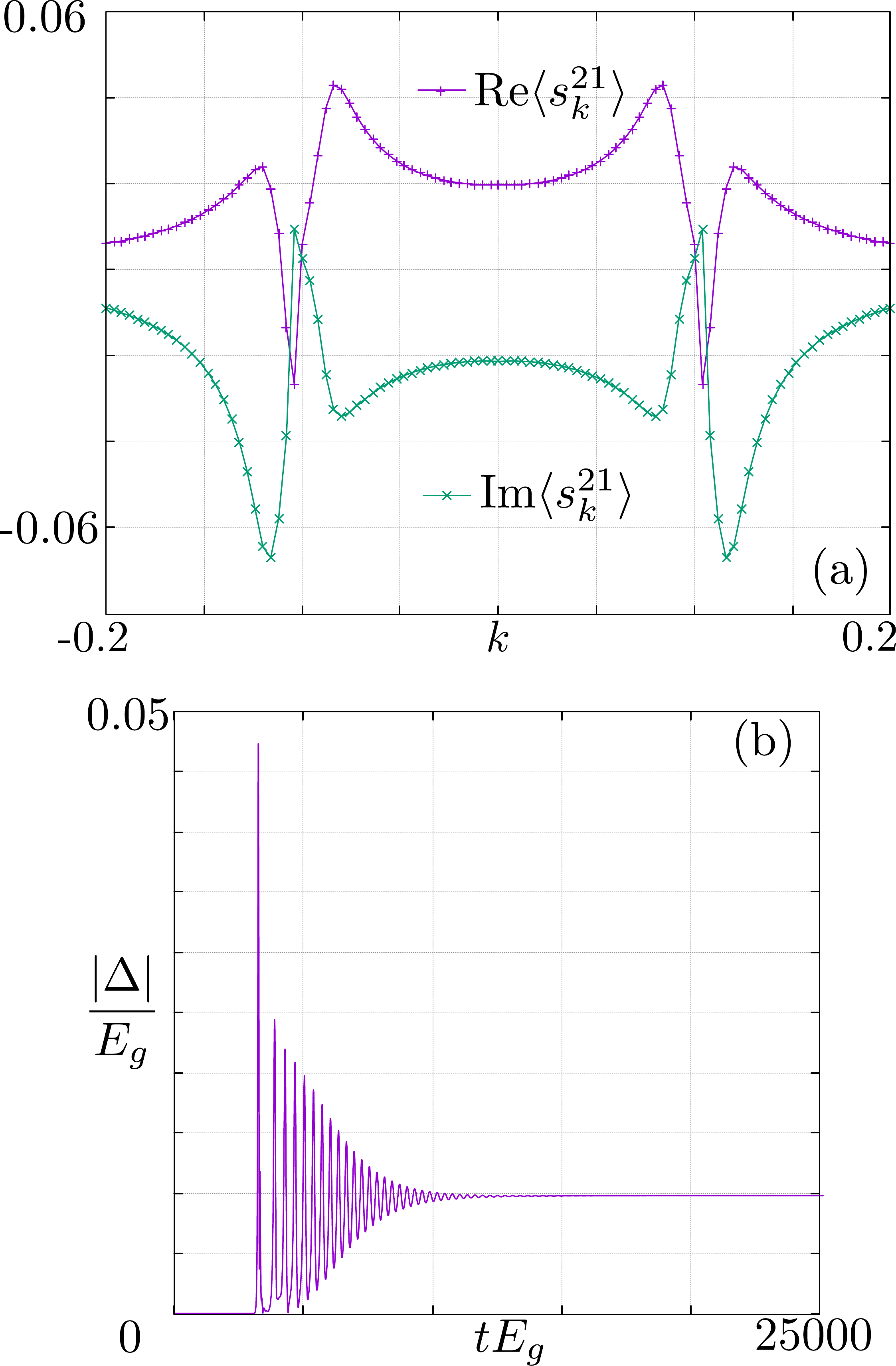}
	\caption{(a) Real and imaginary part of the anomalous interband correlation $ \braket{s^{21}_k} $ in the stationary state. (b) Modulus of the superconducting gap $ \Delta $ as a function of time. Parameters: $ A_1=8E_g $, $ A_2=23E_g $, $ \delta_k=0.2 $, $ \Gamma^{\tmop{ph}}=0.01E_g $, $ \Omega=0.25E_g $ and $ V=5E_g $. Here the interband dissipation is switched on and has the same intensity of the intraband one ($ \Gamma^{\tmop{rr}}=\Gamma^{\tmop{ph}} $). }
	\label{fig:s21delta}
\end{figure}

In this section, we focus on the full system of coupled equation described in App.~\ref{app:eom}. The superconducting order parameter $ \Delta $, which derives from the mean field approximation performed in the density-density interaction Hamiltonian in Eq.~\eqref{eq:Hsys}, is given by:
\begin{equation}\label{eq:delta}
\Delta^*=\dfrac{V}{N}\sum_k \braket{s^{21}_k} \xrightarrow{N\to\infty} \dfrac{V}{2\pi}\int_{-\delta_k}^{\delta_k} \braket{s^{21}_k}  dk 
\end{equation}
Note that we do not need any correction to this formula to achieve a non-zero $ \Delta $ in the stationary state reached by the system. Conversely, in the derivation presented in Ref.~\cite{Goldstein_PRB15}, this is a necessary condition.\\
We start with the initial conditions used in the previous sections. In addition, we set $ \braket{s^{21}_k(t=0)}=10^{-20}(1+i) $. Note that the time evolution is independent of this value: $ \Delta $, as a function of $ t $, always approaches a value numerically compatible with zero before having a jump and stabilizing to its stationary value (see  Figure~\ref{fig:s21delta} (b), where the intensity of the repulsive interaction is set to $ V=5E_g $).\\
\begin{figure}
	\centering
	\includegraphics[width=1\linewidth]{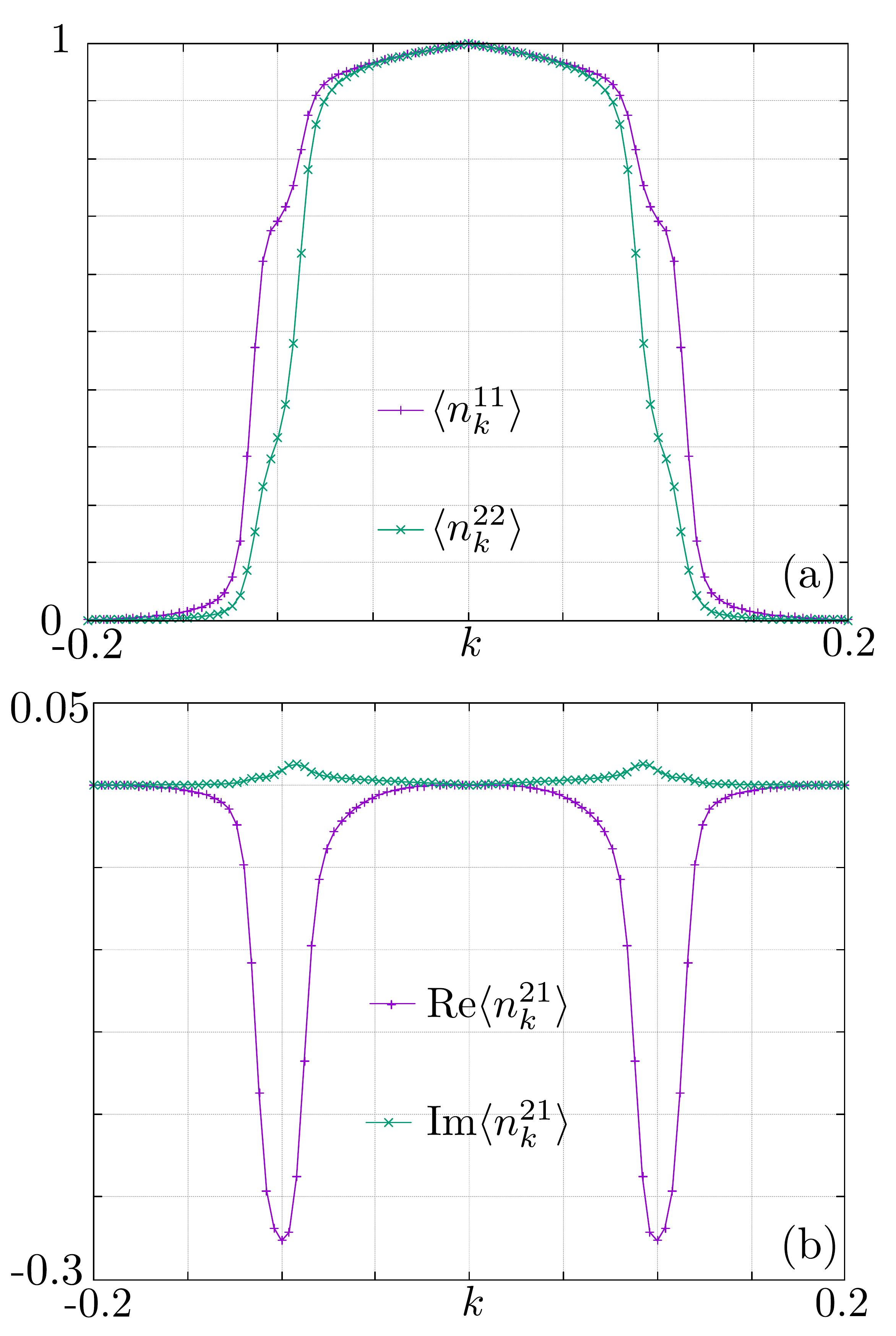}
	\caption{(a) Occupation number of valence $ \braket{n^{11}_k} $ and conduction $ \braket{n^{22}_k} $ bands in the stationary state. (b) Real and imaginary part of the interband correlation $ \braket{n^{21}_k} $ in the stationary state. Parameters: $ A_1=8E_g $, $ A_2=23E_g $, $ \Gamma^{\tmop{ph}}=0.01E_g $, $ \Omega=0.25E_g $ and $ V=5E_g $. Here the interband dissipation is switched on and has the same intensity of the intraband one ($ \Gamma^{\tmop{rr}}=\Gamma^{\tmop{ph}} $). The momentum cutoff is such that $ |k|\leq\delta_k=0.2 $. }
	\label{fig:ndelta}
\end{figure}
We observe in Figure~\ref{fig:ndelta} that in the stationary state the populations are smoothed and do not resemble a box function. Furthermore the interband correlations are not zero in the region where the populations of the valence and conduction bands become slightly different. In this region, where both the bands are partially filled, we find that $ \braket{n_k^{21}}\neq0 $. Then, since the laser light cancels out the energy difference between the valence and conduction bands, the electrons occupy with almost the same probability the two bands. This is also the region where the modulus of the anomalous interband correlator has its maxima (see Fig.\ref{fig:s21delta} (a)). Note that $ \braket{s^{21}_k}\neq 0 $ in the whole momentum region considered, but tends to zero near the boundaries. 

%


\begin{thebibliography}{38}
\bibitem [{Not()}]{Note_gap}%
	\BibitemOpen
	\bibinfo {note} {We  imagine the two bands to be separated  outside the  region in momentum space that we consider. They can, for instance,  bend down.}\BibitemShut {Stop}
\end{thebibliography}
\end{document}